\documentclass{webofc}

\RequirePackage{orcidlink}
\usepackage[varg]{txfonts}   
\usepackage{hyperref}
\usepackage{url}
\usepackage{listings}
\usepackage[symbol]{footmisc}
\hypersetup{colorlinks=true,citecolor=blue,urlcolor=blue,linkcolor=blue}
\begin{document}
\title{Using Containers to Speed Up Development, to Run Integration
Tests and to Teach About Distributed Systems}
%
\author{
        \firstname{Marco} \lastname{Mambelli}\inst{1}\orcidlink{0000-0002-9489-2681}\fnsep\thanks{\email{marcom@fnal.gov}} \and
        \firstname{Bruno} \lastname{Moreira Coimbra}\inst{1}\orcidlink{0009-0002-2797-8706} \and
        \firstname{Namratha} \lastname{Urs}\inst{1}\orcidlink{0000-0001-6217-6376} \and
        \firstname{Ilya} \lastname{Baburashvili}\inst{1}
}
\institute{
    Fermi National Accelerator Laboratory,
    PO Box 500, Batavia IL 60510-5011
          }
\abstract{
GlideinWMS is a workload manager provisioning resources for many experiments, including CMS and DUNE. The software is distributed both as native packages and specialized production containers. Following an approach used in other communities like web development, we built our workspaces, system-like containers to ease development and testing.
Developers can change the source tree or check out a different branch and quickly reconfigure the services to see the effect of their changes.
In this paper, we will talk about what differentiates workspaces from other containers.
We will describe our base system, composed of three containers: a one-node cluster including a compute element and a batch system, a GlideinWMS Factory controlling pilot jobs, and a scheduler and Frontend to submit jobs and provision resources. Additional containers can be used for optional components. This system can easily run on a laptop, and we will share our evaluation of different container runtimes, with an eye for ease of use and performance.
Finally, we will talk about our experience as developers and with students.
The GlideinWMS workspaces are easily integrated with IDEs like VS Code, simplifying debugging and allowing development and testing of the system even when offline.
They simplified the training and onboarding of new team members and summer interns.
And they were useful in workshops where students could have first-hand experience with the mechanisms and components that, in production, run millions of jobs.
}
\maketitle
%


\section{GlideinWMS and HEPCloud}
\label{pilot-wms}
\begin{figure}[htpb]
\centering
\includegraphics[width=8cm,clip]{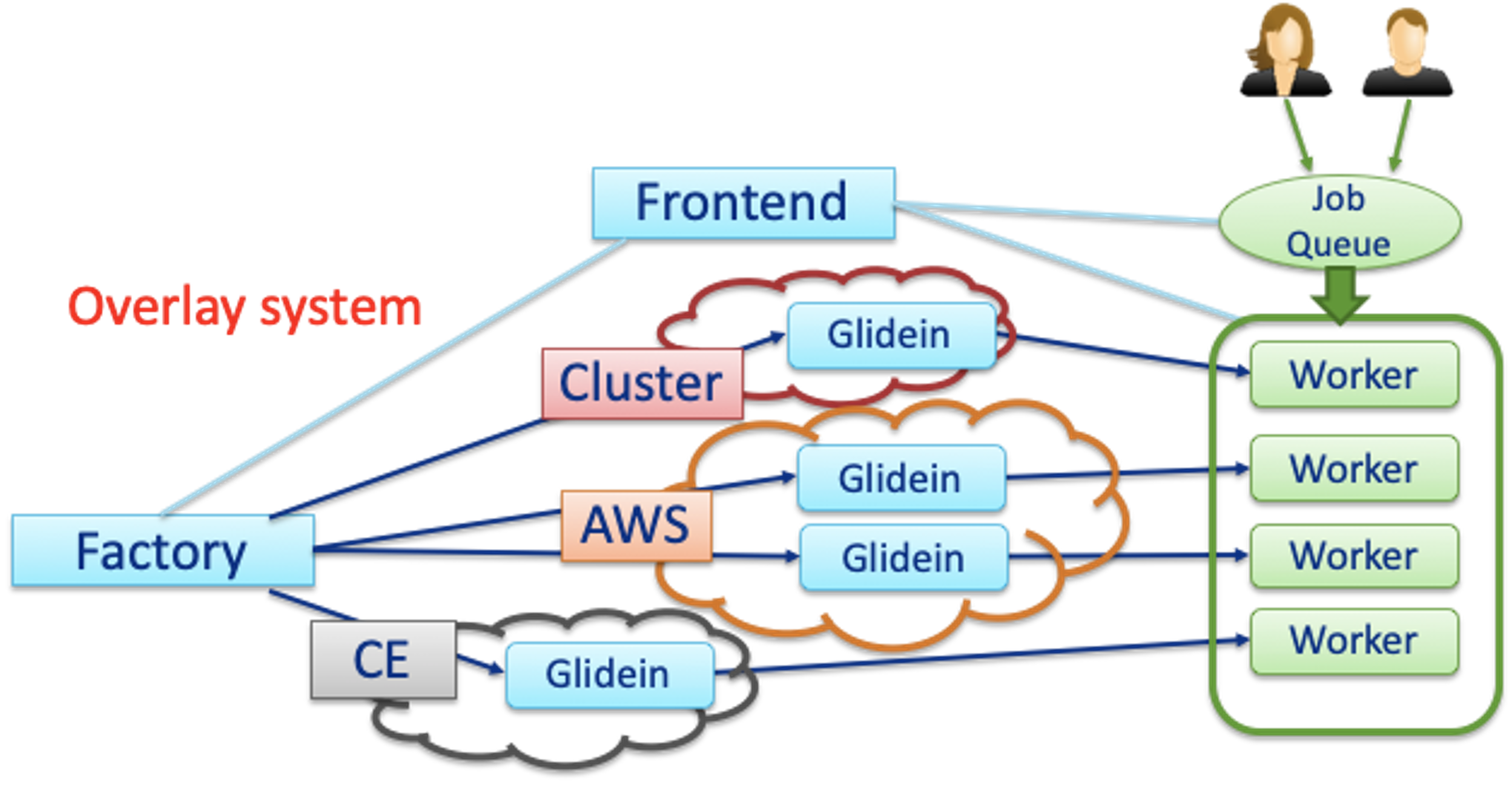}
\caption{GlideinWMS system. GlideinWMS components are in blue, the User Pool is in Green, and the computing resources are in other colors.}
\label{fig-gwms}       
\end{figure}

GlideinWMS~\cite{glideinwms,gwms-sw} (GWMS) is a pilot and pressure-based Workload Management System (WMS) provisioning computing resources in a distributed environment. HEPCloud~\cite{hepcloud} is also a pilot-based WMS, but thanks to its Decision Engine~\cite{hepcloud-de-sw}, it can use more complex resource-provisioning strategies. Their users can request one or more customized elastic HTCondor \cite{htcondor} clusters, User Pools, in green in figure~\ref{fig-gwms}, where the users run their computations.
GlideinWMS provisions the clusters by sending Glideins to a variety of computing resources, also called pilot jobs, to distinguish them from the scientific computations, the user jobs.
GlideinWMS has been and is used at scale in production for more than 10 years by many collaborations, including the Compact Muon Solenoid (CMS) experiment, many Fermilab experiments, and the Open Science Grid (OSG). 
Most scientists do not use GlideinWMS directly or the clusters it provides, instead, they interact with the various tools or portals like CRAB, JobSub, or OSG-Connect, provided by the scientific collaborations.

The Glidein, the pilot job, is the key component in GlideinWMS. It is a program sent to many resources that match the preliminary requirements of the user jobs, to test and set up each computing resource to run the user jobs. 
It manages credentials, provides monitoring and audit information, can auto-detect and report node resources like CPU cores, memory, disk, and GPUs, and can install tools like a container runtime or a distributed file system.
It finally joins the User Pool to run one or more user jobs, in parallel and in sequence, depending on the needs and availability.

The Factory and clients like the Frontend or HEPCloud's Decision Engine complete the GlideinWMS system. For this paper, we will consider a system with one Frontend, one Factory, and their Glideins. Actual deployments may include multiple clients, differing in how they calculate the requests for the Factory, and multiple Factories, providing a redundant distributed system.

The Factory is in charge of submitting Glideins to the different Compute Entrypoints (CEs). 
It knows how to reach each computing resource, which collaborations are supposedly supported, which protocols and authentication methods are supported, and if there are throttling requirements. It submits Glideins, maintaining the pressure on each CE, i.e. number of queued and running Glideins, requested by its clients.
The Factory monitors the Glideins and hosts a secure mailbox to exchange requests and status messages with the clients.

The Frontend and other clients are aware of the users' requests and the running and queued Glideins that can be used for those requests, they receive resource status information from the Factories, and they use heuristics to update the requests to the Factories so that all the user jobs can run promptly, all limits and policies are respected, and no resources are wasted.
The Frontend is generally operated by the scientific experiments or on their behalf and implements their policies for resource provisioning and job priorities.

\begin{figure}[h]
\centering
\includegraphics[width=8cm,clip]{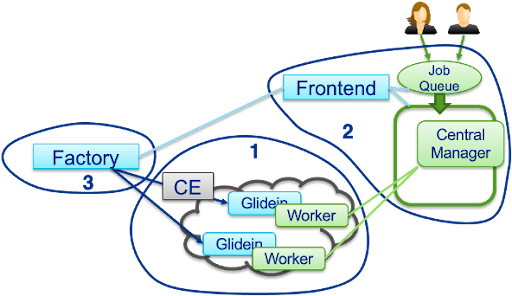}
\caption{A minimal GlidinWMS deployment.}
\label{fig-testbed}       
\end{figure}

A minimal deployment of GlideinWMS constitutes a Factory, a Frontend, a CE, and a virtual cluster using the Glideins as Execution Points.
Figure~\ref{fig-testbed} shows how this can be deployed on three hosts:
a one-node cluster to run the Glideins, with an HTCondor-CE and HTCondor as batch system; a Frontend node also dubbing as Access Point, submit host, and  Central Manager of the User Pool; and a Factory.

\section{Using Containers}
\label{sec-bg}

Containers, as a virtualization technology, have become ubiquitous in cloud native software development and deployment workflows over the past decade. Using virtualization at the operating system level, containers enable multiple isolated environments for applications within a single host.
In contrast to virtual machines, containers are lightweight since they share the same Linux kernel and sometimes more, while virtual machines emulate the whole Operating System and sometimes even the hardware. 
However, thanks to the kernel \emph{cgroups} and \emph{namespaces}, containers can still manage dynamic allocations and provide isolation of processes, files, network, and users.
Containers are very portable since everything needed to successfully run an application anywhere – the code and its dependencies, together with sections of the operating system – is packaged (aka containerized) as a single unit. 
This results in quick, efficient, and effortless deployment and management of applications without having to worry about dependencies and interactions with other installations. 

Containerization has long existed before most modern-day solutions such as Docker~\cite{docker}, Podman~\cite{podman}, and Apptainer~\cite{apptainer}. Docker has accelerated container adoption with its user-friendly platform. At Fermilab, we prefer Podman, a free open-source alternative, mainly for licensing reasons. Apptainer is the platform used to run containers in the Glideins because it is designed for distributed computing: it has compact single-file images, runs as a regular unprivileged application, and minimizes overhead, making nesting easier.
All of these include containers' runtime support and tools to build them. Frequently, there exists an integrated environment, monitoring tools, and a graphical user interface (GUI); all of which make it easier to use containers.

Containers are instantiated starting from a blueprint, the (container) image, one or a set of self-contained, static files that encapsulate all the necessary components of the container, including code, libraries, and configurations. 
A running container, managed by its runtime support, also includes a context and the ability to interact with it.
Images are normally defined using a "recipe" file, e.g., the \lstinline|Dockerfile|, which includes all the instructions to build an image: where to start from, what to install, copy over, or configure. Additionally, containers have orchestration solutions to easily manage and coordinate services running in multiple containers on one or more host nodes. \emph{Docker}, or \emph{Podman}, \emph{Compose} can create volumes and networks and start multiple services with a single command, as directed by its YAML configuration file, the \lstinline|compose| file. Kubernetes and Red Hat OKD can manage and automatically scale multiple instances of many containers.

Traditionally, the GlideinWMS software has been distributed as a tarball or as a native RPM (RPM Package Manager) package, which is installed on production or development hosts. 
The development and testing of the GlideinWMS framework have been mainly carried out on virtual machines hosted in FermiCloud~\cite{Timm_2014}, an infrastructure-as-a-service (IaaS) private cloud deployed at Fermilab using OpenStack~\cite{openstack}.

The normal GlideinWMS software development lifecycle (SDLC) includes active development and a couple of parallel releases, each supporting multiple OS versions, all at the same time.  
This requires many VMs, because a minimal system deployment includes three hosts: a Frontend, a Factory, and a Compute Entrypoint (CE), as described in section \ref{pilot-wms}. The CEs can sometimes be shared by multiple deployments, but not the Factory and Frontend.
FermiCloud always provided several VMs dedicated to GlideinWMS testing and development, some shared across the team and some private for individual developers. However, the number of VMs grew even more during the summer terms due to facilitating workshops and interns collaborating on the GlideinWMS project, which resulted in poor utilization of FermiCloud resources. Furthermore, there were operational overheads to manage the VMs: requesting and renewing host certificates, adapting Puppet configurations, configuring the firewall, extra training requirements, etc. 

A preliminary effort to containerize GlideinWMS resulted from the collaboration with the OSG Consortium, where the members were interested in containers to simplify their production operations of Frontends at first and then of the Factories. Taking inspiration from Web developers like the ones running the Alnoda Hub~\cite{alnoda}, we borrowed the name \emph{workspaces} and designed containers that are similar to virtual machines,  to facilitate quick and easy deployment of the GlideinWMS software, focused towards our development and testing goals.
The workspaces differ from production containers, which are typically microservice-focused, each running a single application. 
Our workspaces are more like Linux hosts, running multiple services and including many tools, which is more conducive for development and testing workflows.


\section{The GlideinWMS Workspaces}
\label{sec-workspaces}
We started with four workspace images: one for each of the nodes in the minimal deployment illustrated in figure~\ref{fig-testbed}: the CE, Frontend, and Factory, and a fourth one, the GWMS workspace, which abstracts the common elements of the other three.
Having a common base allowed us to localize the customizations for the different platforms in that image and parameterize the other three.
We wanted to support multiple RHEL-based OSes, such as Alma Linux 9 with Python 3.9 and Scientific Linux 7 with Python 3.6, on both the AMD/Intel and ARM architectures.

The \textbf{\emph{gwms-workspace}} image comes in three flavors: Alma Linux 9 and Alma Linux 8 with Python 3.9,  and Scientific Linux 7 with Python 3.6, all supporting both \emph{x86\_64} and \emph{aarch64} architectures.
All images start from the official Alma Linux or Scientific Linux image and add some customizations, the OSG RPM repositories, HTCondor, common software packages, and development tools.
The first difficulty was to support multiple services, as in the production hosts. Since containers are designed to run a single service, \lstinline|systemd|~\cite{systemd} does not work in containers. To work around this, we installed \lstinline|supervisord| and used an emulation of the \lstinline|systemctl| command that allows us to use our documented commands to start and stop the GlideinWMS services.

The three additional development blueprints are parametric, i.e., they can extend any of the \lstinline|gwms-workspace| images for the different platforms, and are each responsible for one component of the GlideinWMS framework:
\begin{enumerate}
    \item \textbf{\emph{ce-workspace}} -- is a minimal yet fully functioning SciToken-authenticated computing resource. 
    The template for this workspace installs and sets up the HTCondor batch system, HTCondor-CE, and a few utility scripts. Additionally, this can be configured to fake more computing cores than the available ones.
    \item \textbf{\emph{factory-workspace}} -- is a GlideinWMS Factory, where the underlying template handles the installation and setup of the Factory, including the download of HTCondor tarballs for the Glideins and a working configuration for the minimal deployment. This blueprint also contains an optional setting to link the GlideinWMS installation to a Git repository, so developers can run using the source code instead of a release.
    \item \textbf{\emph{frontend-workspace}} -- is a GlideinWMS Frontend and the HTCondor Access Point (AP) and Central Manager (CM) for the User Pool. The template installs and sets up both the Frontend and the User Pool, and downloads some utility scripts, including the one to link the Git repository, some to refresh credentials, and one to submit jobs to run a smoke test.
\end{enumerate}

To have a fully working test environment, we need the containerized hosts to talk to each other. We opted to use \emph{Docker}, or \emph{Podman}, \emph{Compose}~\cite{dcompose} to orchestrate the containers, which is a simple and easy-to-deploy option, without the need for scaling and multi-host features offered by other solutions.
A \lstinline|compose.yml| file describes the configuration for the deployment of three hosts with the \lstinline|ce-workspace|, \lstinline|factory-workspace|, and \lstinline|frontend-workspace| services (containers). The orchestrator provides a bridged network for the containers with name resolution for the fictitious \emph{glideinwms.org} domain. It also mounts shared volumes on all hosts to share secrets or code. A startup script generates self-signed host certificates from the included trusted CA certificate.

This setup satisfies the basic needs for development, testing, and training. We added a few extensions to cover more use cases.
A parameterized \emph{Compose} configuration allows exposing containers' ports to interact with outside components, e.g., real computing resources not available as containers, like AWS or HPC centers, or GlideinWMS services deployed on hosts or VMs.
To test new releases, we defined an Integration Test-Bed (ITB), the \textbf{\emph{testbed-workspace}}, using an automated deployment script and starting from a new bare-bone image with little more than a minimal OS installation, so that the installation procedure and RPM package dependencies could also be tested.
Finally, we added \textbf{\emph{build-workspace}}, which provides a container to build and serve GlideinWMS releases. Using the GlideinWMS \lstinline|ReleaseManager| tool (included in our software) and RPM tools, this workspace can build release packages starting from any Git reference, local or on GitHub. These can be served by the local YUM server or can be built and distributed by OSG, via its repositories and Koji server. The \emph{build-workspace} can also be included in the ITB to package, install, and test new code automatically.

A note about the decision to support multiple architectures: many developers on our team use Apple Silicon Macs, based on \emph{aarch64} (ARM) processors. Thanks to VMs, it is possible to run \emph{x86\_64} containers on them, and we tested different options using QEMU or Apple's Rosetta 2, but all bring a considerable performance loss, of 20\% or more.
So we decided to add support for the \emph{aarch64} architecture, also called \emph{linux/arm64}. 
This required some modifications to the image description files and a more elaborate build process described in section~\ref{sec-ci-cd-github}, but the resulting use of images, thanks to container manifests, is completely transparent to users, who can simply pull or run the containers while the runtime support ``magically'' chooses the correct architecture.


\section {Continuous Integration/Continuous Deployment}
\label{sec-ci-cd}

For a long time, GlideinWMS has used scripts and GitHub actions and workflows~\cite{ghaction} to automate code testing and the release process.
It was natural to design and develop similar workflows to build workspaces, thereby simplifying the execution of Continuous Integration/Continuous Deployment (CI/CD) pipelines in GlideinWMS.

\subsection {Workspaces CI/CD using GitHub and Docker Hub}
\label{sec-ci-cd-github}
\begin{figure}[htpb]
\centering
\includegraphics[width=8cm,clip]{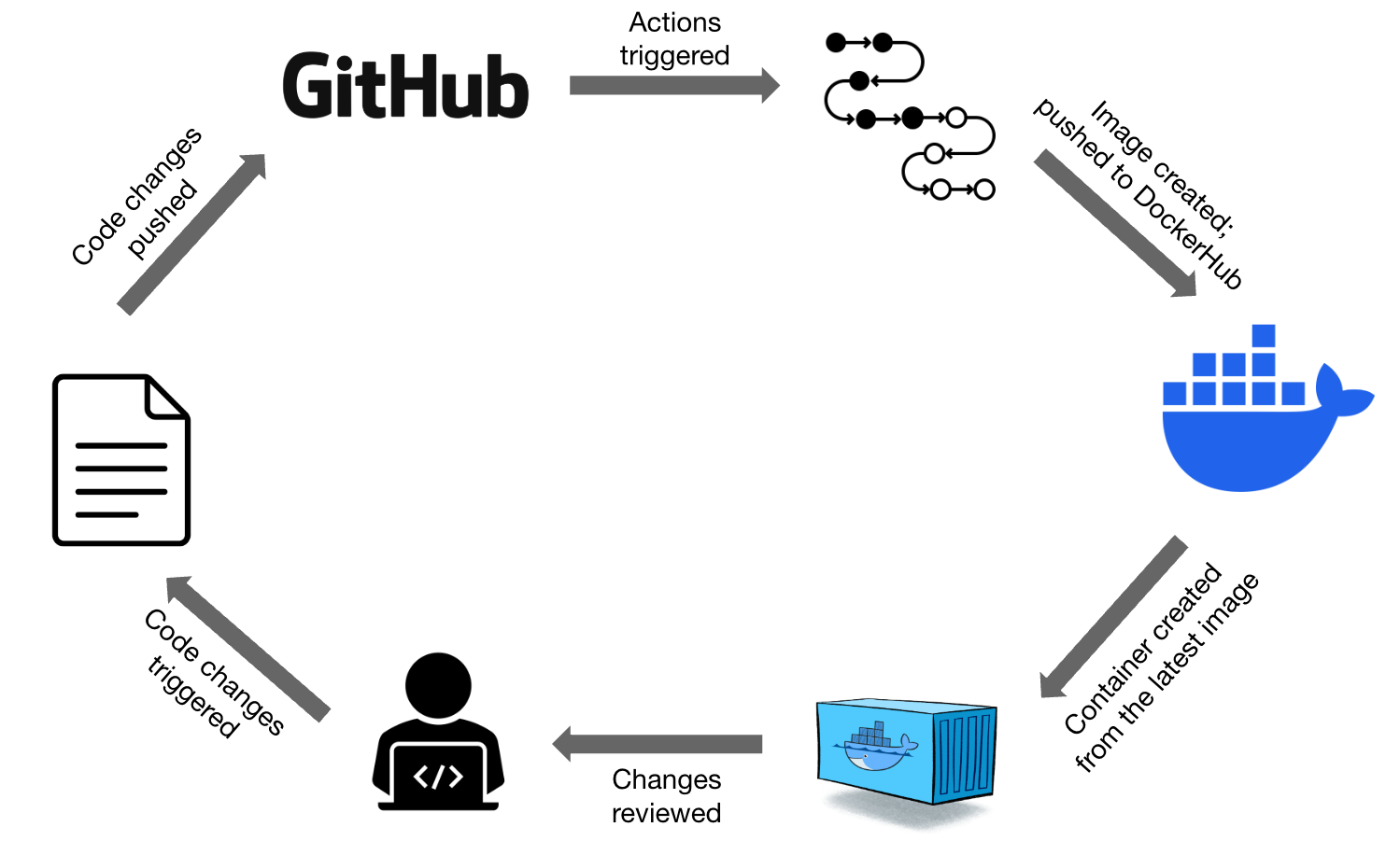}
\caption{Steps of the CI/CD loop for GlideinWMS Workspaces.}
\label{fig-work-cicd}       
\end{figure}

We added GitHub workflows to build workspace images for all the supported platforms and to push them to Docker Hub~\cite{dockerhub}, an image registry.
These images are pulled to run GlideinWMS workspace containers, and the push of changes to the containers in the Git repository triggers the build of new images.
This creates the CI/CD loop shown in figure~\ref{fig-work-cicd}.
The workspaces' GitHub workflow is parametric and works for multiple OS platforms.
It uses Docker Buildx~\cite{dbuildx} to provide manifests and multi-architecture images (\emph{linux/amd64} and \emph{linux/arm64}).
Besides the automatic invocation mentioned above, it can also be triggered via \emph{dispatch}, using GitHub's Web or CLI interfaces. 
The availability of updated multi-platform and multi-architecture images on a well-known public registry, such as Docker Hub, is very convenient to make the workspaces available to other groups or for training events.

\subsection {GlideinWMS CI/CD using Testbed Workspaces}
\label{gwms-cicd-testbed-workspaces}
We are using the \lstinline|testbed-workspace| to test new releases with the addition of scripts to simplify and automate most of the process.
Future work will automate this test and add it to the GlideinWMS CI workflows on GitHub.
This is not trivial because the test requires the tester to authenticate on a Web portal to obtain the SciToken required to submit Glideins to the CE~\cite{gwms-tokens-chep23}.
Fermilab Managed Token service allows setting up a host that can request SciTokens on behalf of the user doing the initial setup. Using the tokens provided, we can fully automate the deployment and testing of new releases on that host. 
The next step will be to devise a similar solution for a dynamic setup like the one on GitHub.


\section{Using GlideinWMS Workspaces for Development, Test, and Build}
\label{sec-dev-test}
The introduction of GlideinWMS workspaces has simplified, sped up, and automated many workflows.
Everyone can run the test system on their laptops and, after downloading the workspace images, they can be offline. Linux is the preferred container platform, but we have instructions to run on Windows using WSL2, and the ARM support allows running natively on M1 Macs, using CoLiMa or Podman. 
You can also run the workspaces from an Integrated Development Environment (IDE) such as Microsoft Visual Studio Code~\cite{gwmswin}.

Developers can quickly spin up a GlideinWMS system, test their changes, switch to different versions, or reproduce an existing setup for troubleshooting.
They don't need to set up and maintain their build server on a Linux host.
The testing of new releases on multiple platforms is almost fully automated.

The project onboarding is very efficient. Even inexperienced interns can run and tweak a complete working system in seconds, a process that used to take weeks. They can change parts of the system and inspect all its details.
We also used the GlideinWMS workspaces for workshops and summer schools, such as the Computational HEP Traineeship Summer School 2024~\cite{hepschool24}.


\section{Conclusion}
\label{sec-conc}
GlideinWMS is a distributed system that can be emulated using at least 3 nodes:
a CE and Cluster, a Frontend and Virtual Cluster, and a Factory.
Workspaces are multi-process containers used to run each of the nodes, and container composition allows the use of a single command to make a dedicated network and run all the containers in the network. We experimented with complex deployments, including multiple clients, adding a node to build and serve new releases, and connecting with external elements. 
Multi-platform container images are distributed via Docker Hub, making the process seamless even on different architectures.
GitHub workflows are used for CI/CD to automate the testing and building of the images that are available on Docker Hub.
The workspaces introduced in this paper have been actively used for development, testing, and training by the GlideinWMS team, other groups within Fermilab, and students, thereby demonstrating their inherent nature of being easily adoptable by others.
Our prior experience in deploying GlideinWMS on virtual machines was extremely valuable in understanding how to effectively design these workspaces and containerize them for rapid deployments, while improving usability.

\section{Acknowledgments}
\label{ack}
The authors’ work was performed using the resources of the Fermi National Accelerator Laboratory (Fermilab), a U.S. Department of Energy, Office of Science, HEP User Facility. Fermilab is managed by Fermi Forward Discovery Group, LLC, acting under Contract No. 89243024CSC000002. 

\bibliography{gwms-hepcloud} 
%
%
%
%

\end{document}